\documentclass[11pt]{article}
\pdfoutput=1

\usepackage{jheppub}

\usepackage{breqn}

\makeatletter
\def\@fpheader{\relax}
\makeatother

\usepackage{amsmath}
\usepackage{graphicx}
\usepackage{amssymb}

\usepackage{multicol}

\usepackage{multicol}
\usepackage{ragged2e}
\usepackage{enumitem}
\usepackage{etoolbox}
\usepackage{booktabs}

\def\S{\Sigma}
\def\s{\sigma}
\def\nn{\nonumber}
\def\p{\partial}
\def\ls{\left[}
\def\rs{\right]}
\def\lc{\left\{}
\def\rc{\right\}}

\newcommand{\bi}{\begin{itemize}}
\newcommand{\ei}{\end{itemize}}

\newcommand{\be}{\begin{eqnarray}}
\newcommand{\ee}{\end{eqnarray}}

\newcommand{\D}{\mathrm{D}}
\newcommand{\Db}{\overline{\mathrm{D}}}

\def\XL{\mathbb{X}_L}
\def\XR{\mathbb{X}_R}
\def\XLb{\overline{\mathbb{X}}_L}

\def\s{\sigma}
\def\nn{\nonumber}
\def\p{\partial}
\def\ls{\left[}
\def\rs{\right]}
\def\lc{\left\{}
\def\rc{\right\}}

\def\p{\partial}

\def\S{\Sigma}
\def\Sb{\overline{\Sigma}}

\def\s{\sigma}

\def\+{{+\!\!\!+}}

\allowbreak 

\allowdisplaybreaks

\title{ 
Superspace Higher Derivative Terms in Two Dimensions 
}

\author[1,2]{Fotis Farakos,}
\emailAdd{farakos@pd.infn.it}

\author[3]{Pavel Ko\v{c}\'{i},}
\emailAdd{pavelkoci@mail.muni.cz}

\author[3]{and Rikard von Unge}
\emailAdd{unge@physics.muni.cz}

\affiliation[1]{Dipartimento di Fisica ÒGalileo GalileiÓ\\
Universit\'{a} di Padova, Via Marzolo 8, 35131 Padova, Italy}

\affiliation[2]{INFN, Sezione di Padova\\
Via Marzolo 8, 35131 Padova, Italy}

\affiliation[3]{ Institute for Theoretical Physics, Masaryk University, \\  611 37 Brno, Czech Republic}

\abstract{
We study $(2,2)$ and $(4,4)$ supersymmetric theories with superspace higher derivatives in two dimensions. 
A characteristic feature of these models is that they have several different vacua, some of which break supersymmetry.  
Depending on the vacuum, the equations of motion describe different propagating degrees of freedom. 
Various examples are presented which illustrate their generic properties. 
As a by-product we see that these new vacua give a dynamical way of generating non-linear realizations. 
In particular, our 2D $(4,4)$ example is the dimensional reduction of a 4D $N=2$ model, 
and gives a new way for the spontaneous breaking of extended supersymmetry. 
}

\preprint{DFPD-2016/TH/08}

\begin{document} 
\maketitle
\flushbottom

\section{Introduction}
Supersymmetric theories often exhibit dualities, 
which are more conveniently studied in superspace \cite{Gates:1983nr}. 
These superspace  dualities come in two different types. 
In most cases of interest, the duality is a {\em true} duality in the sense that it changes the description of the underlying physics. 
The best known example is T-duality in two-dimensional sigma models where the target space geometry is changed \cite{
Hitchin:1986ea,
Hull:1985pq,
Rocek:1991ps,
Hull:1991uw,
Grisaru:1997ep,
Maes:2006bm}. 
One may also think of the duality between a chiral superfield and a real linear superfield (i.e. a tensor multiplet) in four dimensions, 
where the scalar field is exchanged with a two form \cite{Siegel:1979ai,Lindstrom:1983rt}. 
This type of duality relies on special properties of the theory at hand and it cannot be performed in every case. 
For example, 
the chiral superfield can be dualized to a real linear superfield only when the theory is independent of the imaginary part of the chiral superfield. 
In the case of T-duality, the duality exists only when isometries are present in the target space geometry. 

A second type of superspace dualities exists though, which can be generically performed. 
This duality in principle does not rely on the special properties of the model under consideration. 
However, it is {\em not} a true duality, 
since the description of the underlying physics is left unchanged. 
The typical example of this type of duality is the duality between chiral superfields and complex linear superfields 
\cite{
Gates:1980az,
Deo:1985ix,
Grisaru:1997hf,
Penati:1997pm}, 
which exists both in four and in two dimensions. It has been shown that performing this duality in a K\"ahler sigma model does not change the target space geometry, 
instead, it just changes the original coordinates into new coordinates which are not related to the complex structure of the K\"ahler geometry \cite{Penati:1998yt}. 
This second type of duality has been used to argue that in supersymmetric models with scalar fields it is enough to consider only chiral superfields (as well as twisted chiral \cite{Gates:1984nk} and semichiral superfields \cite{Buscher:1987uw} in two dimensions), because any model of complex linear superfields can be dualized to a chiral model.

However, this viewpoint recently changed when it was shown that models of complex linear superfields 
can break supersymmetry in an entirely new fashion \cite{Farakos:2013zsa,Farakos:2014iwa,Farakos:2015vba}, 
and violate the chiral-complex linear duality. 
The simplest model is described by the Lagrangian 
\begin{align}
\label{Theory}
{\cal L} = - \int d^4\theta \;\overline\S\S 
+ \frac{1}{8f^2}\int d^4\theta \; \D^\alpha\S \, \D_\alpha\S \, 
\overline{\D}^{\dot\alpha} \overline\S \, \overline{\D}_{\dot\alpha}\overline\S\, .
\end{align}
The higher derivative term in \eqref{Theory} introduces a nontrivial potential for the auxiliary field $F = \D^2\S|$. 
The equations of motion for $F$ now have several solutions with different physics. 
In addition to the trivial supersymmetry preserving solution $F=0$, 
there is also  a new solution with $F \neq 0$ 
where supersymmetry is spontaneously broken (and non-linearly realized). 
In this background the previously auxiliary fermion $\lambda_\alpha  = \D_\alpha\S |$ 
becomes propagating and plays the role of the massless fermion accompanying supersymmetry breaking: the goldstino. 
Therefore, in distinction to the theory of chiral superfields where the supersymmetry breaking is introduced through a linear superpotential, 
here the supersymmetry breaking is introduced by a superspace higher derivative term. 
Furthermore, instead of the physical fermion becoming the goldstino, as is the case for the chiral superfield, 
here this role is played by a previously auxiliary fermion. 
Therefore, in the vacua where supersymmetry is broken, there is an additional complex fermion propagating. 
Since the number of degrees of freedom is always preserved in all known duality procedures, 
this model explicitly breaks the duality between chiral and complex linear superfields \cite{Farakos:2015vba}.

In this article we extend the study of superspace higher derivatives to 
models in two dimensions 
with $(2,2)$ and $(4,4)$ supersymmetry.
As we will see, these models have new vacuum solutions, compared to the free theories, 
which propagate different degrees of freedom and manifestly break the duality. 
Our motivation for this study is twofold: 
Firstly, we want to investigate if this mechanism is specific for 4D $N=1$ supersymmetry, or if it is a generic property of supersymmetric theories. 
Our  findings indicate that the structure of supersymmetric theories generically will allow for such a mechanism to occur, 
and we give various examples in two dimensions. 
We also show how this mechanism will work in theories with 8 supercharges (4D $N=2$ supersymmetry). 
Secondly, 
there are many reasons to expect the two-dimensional case to be richer and more exciting than the four dimensional one. In two dimensions Lorentz invariance is much less restrictive since it acts only by rescaling the various spinors. 
In addition, 
the auxiliary complex vector $P_{\alpha\dot\alpha} = \ls \D_\alpha , \overline{\D}_{\dot\alpha} \rs\S |$ of the 
four dimensional complex linear superfield reduces in two dimensions to a complex vector and two additional scalar auxiliary 
fields which will introduce many new solutions to the equations of motion. 
In two dimensions there are also several other inequivalent representations with the same on-shell content that can play a role.
For example we find that there are vacua where the goldstino will reside in a nilpotent {\em  twisted} chiral superfield, 
or that some vacua give rise to chiral bosons \cite{Siegel:1983es,Gates:1987sy,Sevrin:2013nca}. 
We also investigate how semi-chiral multiplets \cite{Buscher:1987uw,Goteman:2012qk,Lindstrom:2014bra} 
behave in the presence of higher derivative terms.

\section{Goldstino multiplets in $\mathbf{(2,2)}$}

In the examples we will study in the following, 
we will see that there exist various backgrounds where supersymmetry is spontaneously broken and non-linearly realized. 
The generic property of broken supersymmetry   
is the existence of Goldstone fermionic modes: the goldstini. 
The $(2,2)$ supersymmetry transformations of these fermions have the property 
\be
\begin{aligned}
\langle \delta G_+ \rangle = f \,  \epsilon_+ \,,  
\\
\langle \delta G_- \rangle = f \, \epsilon_- \,, 
\end{aligned} 
\ee
where $f$ is the supersymmetry breaking scale, 
which in 2D has mass dimension 1. 
To be prepared to identify these backgrounds in the superspace setup, 
we dedicate this section to the superspace formulation of  $(2,2)$ supersymmetry breaking, 
and the different ways of embedding the goldstini inside superfields. 
Before we start our discussion, 
let us mention that the $(2,2)$ superspace  algebra we use is 
\begin{equation}
\begin{aligned}
\{ \D_- , \Db_- \} &= i \p_{=}  \,, 
\\
\{ \D_+ , \Db_+ \} &= i \p_{\+} \,, 
\end{aligned} 
\end{equation} 
while all other anti-commutators vanish, i.e. there are no central charges. 
Definitions of the component fields and our conventions can be found in the appendix.

We start by describing the embedding of the goldstini in a chiral superfield 
\be
\Db_-  X = 0 = \Db_+ X \,.  
\ee 
Mimicking the 4D theory \cite{Rocek:1978nb,Lindstrom:1979kq,Casalbuoni:1988xh}, 
once we impose the constraint 
$X^2 = 0$, 
the scalar lowest component of $X$ is removed from the spectrum, 
and we have in chiral coordinates 
\be
X = \frac{G_- G_+}{\cal F} + \theta_+ G_- + \theta_- G_+ + \theta_+ \theta_- {\cal F} \,. 
\ee
Here $\cal F$ is the auxiliary field which gets a {\it vev} and breaks supersymmetry. 
In the following sections we will need the superspace equations of motion for the supersymmetry breaking sector. 
To derive them it is convenient to write a superspace Lagrangian formulation where the constraint is imposed by a Lagrange multiplier. 
The simplest model is described by the superspace Lagrangian 
\be
{\cal L} = \int d^4 \theta X \overline X - \lc \int d^2 \theta  \left(  f  X + {\cal M} \, X^2 \right)+ c.c. \rc  \,, 
\ee
where $X$  is for the moment a standard chiral superfield and ${\cal M}$ is the chiral Lagrange multiplier superfield. 
The variation of the chiral superfields  $\cal M$ and $X$ yields  
\be
\label{X1}
X^2 &=& 0 \,, 
\\
\label{X2}
\overline{\D}_+  \overline{\D}_- \overline X &=& - f - 2 \, {\cal M}\, X \,. 
\ee
These superspace equations correspond to the equations of motion for the goldstini $G_-$ and $G_+$. 
Equation \eqref{X2} also gives a {\it vev} to $\cal F$.

In 2D there is an additional scalar multiplet, 
which can describe the goldstini in a simple fashion.  
We can have a {\it twisted} chiral superfield 
\be
\label{twistednilp}
\Db_+  Y = 0 = \D_- Y \,. 
\ee 
Similarly as for the chiral, 
once we impose the constraint 
$Y^2 = 0 $ 
the scalar lowest component of $Y$  is removed from the spectrum and we have in chiral coordinates 
\be
Y = \frac{ G_+ \overline G_-}{\cal F} + \overline \theta_+ \overline G_- +  \theta_- G_+ + \theta_- \overline \theta_+  {\cal F}  \,. 
\ee 
To find the superspace equations of motion for the nilpotent 
twisted chiral superfield we have to impose the nilpotency condition $Y^2=0$ by a Lagrange multiplier. 
The simplest Lagrangian reads 
\be
{\cal L} = \int d^4 \theta \, Y \overline Y - \lc \int d^2 \theta^T  \left(  f \, Y + {\cal N} \, Y^2 \right)+ c.c. \rc  \,, 
\ee
where $Y$ is a twisted chiral superfield, but otherwise unconstrained, and  ${\cal N}$ is a twisted chiral Lagrange multiplier superfield. 
The superspace equations of motion give 
\be
\label{Y1}
Y^2 &=& 0\,,
\\
\label{Y2}
\D_-  \overline{\D}_+ \overline Y  &=& -  f - 2 \, {\cal N}\, Y \, .  
\ee
These superspace equations correspond to the equations of motion for the goldstini $\overline G_-$ and $G_+$.

An alternative formulation for the description of the $(2,2)$ goldstini can be provided by 
direct analogy to  the 4D spinor goldstino superfield \cite{Ivanov:1978mx,Ivanov:1982bpa,Samuel:1982uh,Kuzenko:2011ti,Bandos:2016xyu}. 
In two dimensions we define the  superfields $\Lambda_+$ and $\Lambda_-$ by the equations 
\be 
\label{SWdef}
\begin{aligned}
\D_+ \Lambda_+ &= 0 = \D_- \Lambda_-  \, , 
\\
\D_- \Lambda_+ &= -\kappa = - \D_+ \Lambda_-  \, , 
\\
\overline{\D}_+\Lambda_\pm & = \frac i\kappa \Lambda_-\partial_{\+}\Lambda_\pm  \, , 
\\
\overline{\D}_-\Lambda_\pm & = -\frac i\kappa\Lambda_+\partial_=\Lambda_\pm  \, .  
\end{aligned}
\ee
Here $\kappa$ is a constant of mass dimension $1$ that will be related to the vacuum expectation value of the auxiliary scalar field. 
By redefining the $\Lambda$ we can always make $\kappa$ to be real. 
The simplest Lagrangian for these superfields is 
\be
{\cal L} =  - \int d^4 \theta \, \Lambda_- \Lambda_+ \overline{\Lambda}_- \overline{\Lambda}_+ \,. 
\ee
It is quite intriguing that one needs both $\Lambda_+$ and $\Lambda_-$ for the correct description of the 2D spinor goldstino superfield.

Following \cite{Samuel:1982uh}, 
we can build chiral or twisted chiral superfields which describe the $(2,2)$  goldstini, 
in terms of the $\Lambda_\alpha$ superfields, as 
\be
\begin{aligned}
X &= \overline{\D}_- \overline{\D}_+ (  \Lambda_- \Lambda_+ \overline{\Lambda}_- \overline{\Lambda}_+  ) \,, 
\\
Y &= \overline{\D}_+ \D_- (  \Lambda_- \Lambda_+  \overline{\Lambda}_- \overline{\Lambda}_+  )\, . 
\end{aligned}
\ee  
Notice that $X^2=0$ and $Y^2=0$. 
We may also construct four different complex linear superfields (in fact they are left or right semichiral superfields but linear combinations of them will in general be only complex linear)
\be
\begin{aligned}
\S_{F-} &= \overline{\D}_-(\Lambda_- \Lambda_+\overline{\Lambda}_+) 
= \Lambda_-\Lambda_+\left(\kappa -\frac i\kappa \overline{\Lambda}_+\partial_=\Lambda_+\right) \,,  
\\
\S_{F+} &= \overline{\D}_+(\Lambda_+\Lambda_-\overline{\Lambda}_-) 
= \Lambda_-\Lambda_+\left(\kappa - \frac i\kappa\overline{\Lambda}_-\partial_{\+}\Lambda_-\right)\,, 
\\
\S_K &= \overline{\D}_-(\overline{\Lambda}_+\overline{\Lambda}_-\Lambda_+) 
=\overline{\Lambda}_- \Lambda_+ \left(\kappa - \frac i\kappa\overline{\Lambda}_+\partial_=\Lambda_+\right) \,,  
\\
\label{build}
\S_L &= \overline{\D}_+(\overline{\Lambda}_+\overline{\Lambda}_-\Lambda_-) 
=\overline{\Lambda}_+ \Lambda_- \left(\kappa - \frac i\kappa\overline{\Lambda}_-\partial_{\+}\Lambda_-\right) \,.  
\end{aligned}
\ee
From here it is clear that $\S_{F-}^2=\S_{F+}^2 = \S_K^2=\S_L^2=0$ as well as $\S_{F-}\S_K=\S_{F-}\S_L = \S_{F+}\S_{K} = \S_{F-}\S_L = \S_{F-}\S_{F+} = 0$ while $\S_K\S_L\neq 0$. 
The nilpotent complex linear superfields in \eqref{build} 
are indexed by the respective scalar auxiliary field that gets a vacuum expectation value (see the appendix for the definition of all components).
We see that $\S_{F\pm}$ both have a nonzero vacuum expectation value for the $F$ auxiliary field whereas $\S_K$ and $\S_L$ have nonzero values for the $K$ and $L$ auxiliary scalars that in four dimensions would be part of the vector auxiliary field. 
It is satisfying to observe that in correspondence to the increased number of auxiliary 
scalar fields for the two-dimensional complex linear superfield 
there is a corresponding increase in the number of ways 
that we may build a complex linear superfield in terms of the spinor goldstino supermultiplet. 
Notice that the same building blocks \eqref{SWdef} can be used to build twisted complex linear superfields.

We will also study models where 2D $(4,4)$ supersymmetry is spontaneously
broken and non-linearly realized. The properties of these models are
equivalent to the 4D $N=2$ \cite{Kuzenko:2011ya,Cribiori:2016hdz}, and we
follow the formulation given in \cite{Cribiori:2016hdz}. We will give more
details in the later sections when we study these models.

\section{Models with complex linear multiplets in $\mathbf{(2,2)}$}

The free model of complex linear superfields has a superspace Lagrangian \cite{Deo:1985ix} 
\begin{equation} 
\label{V01}
{\cal L} =  \int d^4 \theta \Big{[} - \alpha \, \overline{\S } \S  - \beta \, \S^2 - \beta \, \overline \S^2  \Big{]} \, ,  
\end{equation} 
which on-shell describes the same propagating modes as the free chiral multiplet. 
The parameters $\alpha$ and $\beta$ can be real without loss of generality, 
and one can have $\beta=0$ while keeping the same propagating degrees of freedom. 
If we set $\alpha =0$ or $\alpha = \pm 2  \beta$, then from the equations of motion one can see that there are no propagating modes. 
In component form the Lagrangian \eqref{V01} for $\beta=0$ and $\alpha=1$ reads (after we eliminate the auxiliary fields)  
\begin{equation} 
\label{V01comp}
{\cal L} =  \p_= A \, \p_{\+} \overline A  +  i \psi_- \p_{\+} \overline{\psi}_- + i \psi_+ \p_{=} \overline{\psi}_+   .  
\end{equation}

In principle, 
by adding self-interaction terms to the Lagrangian \eqref{V01}, 
it is expected that the number of propagating degrees of freedom will not change. 
As has been pointed out in \cite{Farakos:2013zsa,Farakos:2014iwa,Farakos:2015vba}, this is not always the case. 
We will see that by adding superspace higher derivatives to the free theory \eqref{V01}, 
it not only becomes self-interacting, 
but it has different propagating degrees of freedom. 
In all our examples we show how the content of propagating modes is altered once we introduce the superspace higher derivative terms in the theory. 
As we will see this gives rise to new vacuum solutions which describe various on-shell theories that are different from the free one 
not only in the sense that the interactions are different, but it is the propagating fields themselves that are different. 
The simplest terms\footnote{In this paper we focus on terms which are proportional to $(\D \S)^4$ but one can find lower dimensional operators that could generate a non-trivial supersymmetry breaking vacuum. However these terms have usually propagating ghosts in the bosonic spectrum and we will not study such cases here.} one can write down are the ones which are manifestly real.  
We have 
\begin{equation}
\label{HDterms11}
\begin{aligned}
\int d^4 \theta \Big{[} 
\D_+ \S \, \D_- \S \,  \Db_+ \overline \S \, \Db_- \overline \S 
\Big{]} & \sim 
- K \overline K L \overline L - (F \overline F)^2  + F \overline F ( K \overline K + L \overline L) \, , 
\\ 
\int d^4 \theta \Big{[} 
\Db_+ \S \, \Db_- \S \,  \D_+ \overline \S \, \D_- \overline \S 
\Big{]} & \sim 
- K \overline K L \overline L \, , 
\\ 
\int d^4 \theta \Big{[} 
\D_+ \S \, \Db_- \S \,  \Db_+ \overline \S \, \D_- \overline \S 
\Big{]} & \sim 
 (L \overline L)^2 - F \overline F L \overline L \, , 
\\ 
\int d^4 \theta \Big{[} 
\Db_+ \S \, \D_- \S \,  \D_+ \overline \S \, \Db_- \overline \S 
\Big{]} & \sim 
(K \overline K)^2 - F \overline F K \overline K \, , 
\end{aligned}
\end{equation} 
where we have also sketched the scalar auxiliary fields contribution. 
We will see that by adding these terms, 
the theory has new vacua, and the propagating degrees of freedom are different. 
In fact it is not merely the presence of the higher derivatives which gives rise to the new degrees of freedom, 
but the fact that the higher derivatives give access to new vacua, which contain different degrees of freedom.

For completeness we give here a list of all the other possible higher derivative terms with similar structure as \eqref{HDterms11} that one could add.  
In this paper we will not study them further, but in each case we list the scalar auxiliary field potential. They are 
\be
\nn
\int d^4 \theta \Big{[} 
\D_+ \S \, \D_- \S \,  \D_+  \Sb \, \D_-  \Sb + c.c.
\Big{]} & \sim & 
 F^2 \overline K  \overline L + \overline F^2  K L \, , 
\\ 
\nn
\int d^4 \theta \Big{[} 
\D_+ \S \, \D_- \S \,  \D_+  \Sb \, \Db_-  \S + c.c.
\Big{]} & \sim    & 
K \overline K ( F L + \overline F \overline L ) \, , 
\\
\nn
 \int d^4 \theta \Big{[} 
\D_+ \S \, \D_- \S \,  \D_+  \Sb \, \Db_-  \Sb + c.c.
\Big{]} & \sim    & 
(K \overline K - F \overline F) (K \overline F + \overline K F) \, ,
\\
\nn
 \int d^4 \theta \Big{[} 
\D_+ \Sb \, \D_- \Sb \,  \D_+  \S \, \Db_-  \S + c.c.
\Big{]} & \sim   & 
- L \overline L (F \overline K + \overline F K) \, ,
\\
\nn
\int d^4 \theta \Big{[} 
\D_+ \S \, \D_- \S \,  \Db_+  \S \, \Db_-  \S + c.c.
\Big{]} & \sim    & 
- L^2 K^2 - \overline L^2 \overline K^2 \, ,
\\
\nn
\int d^4 \theta \Big{[} 
\D_+ \S \, \D_- \S \,  \Db_+  \S \, \Db_- \overline \S + c.c.
\Big{]} & \sim    & 
(F \overline F - K \overline K) ( K L + \overline K \overline L )  \, ,
\\ 
\nn
\int d^4 \theta \Big{[} 
\D_+ \S \, \D_- \S \,  \Db_+  \overline \S \, \Db_-  \S + c.c.
\Big{]} & \sim    & 
(F \overline F - L \overline L) ( K L + \overline K \overline L )  \, , 
\\ 
\nn
\int d^4 \theta \Big{[} 
\D_+ \S \, \D_- \overline \S \,  \Db_+  \S \, \Db_-  \S + c.c.
\Big{]} & \sim    & 
- L \overline L ( L K +  \overline L \overline K ) \, ,
\\ 
\int d^4 \theta \Big{[} 
\D_+ \overline \S \, \D_- \S \,  \Db_+ \S \, \Db_-  \S + c.c.
\Big{]} & \sim    & 
- K \overline K ( L K +  \overline L \overline K ) \, . 
\ee

In 2D, we may investigate the same effect using twisted complex linear superfields, defined as 
\begin{equation} 
\Db_+ \D_-  \Omega = 0 \, . 
\end{equation}
All the results are completely equivalent to the analysis performed with the complex linear superfields so we do not investigate them further. However, it might be interesting to study models containing both complex linear and twisted complex linear fields in this context. We leave this for the future.

\subsection{The model from 4D N=1}

Let us dimensionally reduce the model  
(\ref{Theory}) 
from four to two dimensions. 
We get
\begin{equation} 
\label{V2}
{\cal L} =  \int d^4 \theta \Big{[} - \overline{\S } \S  
-  \frac{1}{2 f^2} \, \D_+ \S \, \D_- \S \,  \Db_+ \overline \S \, \Db_- \overline \S \Big{]} \, . 
\end{equation} 
To find the different vacua we focus for the moment on the bosonic sector
\begin{equation}
\begin{aligned}
{\cal L}_{bos.}  =&\ \p_= A \, \p_{\+} \overline A - P_{\+} \overline{P}_= - \overline{P}_{\+} P_=  -F \overline F + K \overline K + L \overline L 
\\
& + \frac{1}{2 f^2} \Big [ P_{\+} P_= \overline{P}_{\+} \overline{P}_=  - \overline K \overline L P_{\+} P_= 
- K L \overline{P}_{\+} \overline{P}_= + F \overline F (\overline{P}_{\+} P_= + P_{\+} \overline{P}_= ) \Big ]  
\\
& + \frac{1}{2 f^2} \Big [  - F \overline F (  K \overline K + L \overline L) +  K \overline K L \overline L + (F \overline F)^2 \Big ]  \, ,
\end{aligned}
\end{equation}
and it is clear that the variational equations for the vector auxiliary fields give 
\begin{equation}
P_{\+} = 0 = P_=  \, . 
\end{equation}
Note however that one could investigate more exotic examples where the vectors have a non-vanishing expectation value, explicitly breaking Lorentz invariance,
but we will not investigate this possibility further here. 
The variation of the auxiliary fields $F$, $K$ and $L$ gives 
\begin{equation}
\begin{aligned}
0 = & \, \overline F \Big[ 1 + \frac{1}{2 f^2} \left(K \overline K +  L \overline L - 2  F \overline F \right) \Big] \, , 
\\
0 = & \, \overline K \Big[ 1 + \frac{1}{2 f^2} \left( L \overline L - F \overline F \right) \Big] \, , 
\\
0 = & \, \overline L \Big[ 1+ \frac{1}{2 f^2} \left( K \overline K -  F \overline F \right) \Big]\, . 
\end{aligned}
\end{equation}
In addition to the trivial solution 
\begin{equation}
F = K = L = 0 \, , 
\end{equation}
the above equations
admit various other solutions, the simplest possibility being 
\begin{equation}
F = f \ , \ K=L=0 \, . \label{V1}
\end{equation} 
One can now easily see that the bosonic sector of the theory in the background \eqref{V1}  is  
\begin{equation}
{\cal L}_{bos.}  = - \frac{f^2}{2} + \p_= A \, \p_{\+} \overline A \, . 
\end{equation} 
The positive vacuum energy implies that supersymmetry is broken.

Now we investigate the fermionic sector in the background \eqref{V1}, 
and we keep only terms quadratic in fermions which are useful to uncover the propagating degrees of freedom. 
We have  
\begin{equation}
{\cal L}_{ferm.}^{(2)}  = i \psi_- \p_{\+} \overline{\psi}_- + i \psi_+ \p_{=} \overline{\psi}_+  
+ \frac i 2 \lambda_+ \p_{=} \overline{\lambda}_+ + \frac i 2 \lambda_- \p_{\+} \overline{\lambda}_-  \, .  
\end{equation} 
Notice that in contrast to the free theory (which is the one we recover when $F=0=K=L$) there are 
additional propagating degrees of freedom: the fermions $\lambda_+$ and $\lambda_-$ . 
From the supersymmetry transformations of these fermions we see that they are the goldstini 
of the broken supersymmetries 
\begin{equation}
\begin{aligned}
\langle \delta \lambda_+ \rangle & = \epsilon_+ f \, , 
\\
\langle \delta \lambda_- \rangle & = \epsilon_- f \, . 
\end{aligned}
\end{equation}
On the other hand the fermions $\psi_+$ and $\psi_-$ are not goldstini, but they make an on-shell $(2,2)$ supermultiplet 
together with the boson $A$.

We can now verify that the solution \eqref{V1} can be lifted to a full solution of the equations of motion. 
The more transparent way to see this is by solving the superspace equations of motion 
from Lagrangian \eqref{V2} 
\begin{equation}
\label{V3}
\S
- \frac{1}{2 f^2} \Big{[} 
\Db_+ \big( \Db_-  \overline \S \, \D_+  \S \, \D_-  \S \big )
- \Db_- \big( \Db_+  \overline \S \,  \D_+  \S \, \D_-  \S \big )
\Big{]}  
= \overline \Phi \, , 
\end{equation}
where now $\Phi$ is a chiral superfield which is the zero-mode of the superspace derivative $\mathrm{D}_\alpha$. 
As a consistency condition arising from \eqref{V3} when acting with $\Db_+  \Db_- $ we see that $\Phi$ satisfies 
\begin{equation}
\label{V6}
\Db_+  \Db_-  \overline \Phi = 0 \, . 
\end{equation} 
The equation \eqref{V3} then has a solution of the form 
\begin{equation}
\S = \overline \Phi + X  \, , 
\end{equation}
where $X$, as we defined it earlier satisfies 
\begin{equation}
\label{V4}
\begin{aligned}
 X^2 &= 0 \, ,  
 \\ 
\D_+ \D_- X  &= f + 2 \, \overline {\cal M} \,  \overline  X \, , 
\end{aligned} 
\end{equation}
and where $ \cal M$ is a chiral Lagrange multiplier. 
Now it is easy to match the results from the component form, 
to the superspace results we just found. 
One has 
\begin{equation} 
A, \psi_+, \psi_-  \in \overline \Phi \, ,
\end{equation}
and 
\begin{equation}
\lambda_- , \lambda_+ \in X \, . 
\end{equation}

It is important at this point to notice that the superspace equations \eqref{V3} admit the simple and exact solution \eqref{V4}, 
because the zero-mode $\Phi$ drops out of the higher derivative part of the equations of motion, 
leaving behind the equation 
\begin{equation}
\label{V5}
X
- \frac{1}{2 f^2} \Big{[} 
\Db_+ \big( \Db_-  \overline X \, \D_+  X \, \D_-  X \big )
- \Db_- \big( \Db_+  \overline X \,  \D_+  X \, \D_-  X \big )
\Big{]}  
= 0 \, , 
\end{equation}
which is solved by \eqref{V4}. 
In this way the equations of motion of the zero-mode $\Phi$ \eqref{V6} 
and the equations of motion of the goldstino superfield $X$ \eqref{V5} decouple, 
and they are solved easily independently. 
In the next sections, 
where we are going to investigate more complicated solutions, 
the condition that the zero-mode drops out of the higher derivative terms will guide us to find exact results.

For completeness we also show how the goldstino is embedded using the 2D version of the Samuel-Wess superfield
$\Lambda_\pm$ (\ref{SWdef}).
If we define
\begin{equation}
\S =  c_+ \S_{F+} + c_- \S_{F-} 
=   c_+ \, \Db_+ \left( \Lambda_+ \Lambda_- \overline \Lambda_+  \right) 
+  c_- \, \Db_- \left( \Lambda_+ \Lambda_- \overline \Lambda_-  \right) \, , 
\end{equation} 
and insert this into Lagrangian \eqref{V2} we find (for example we choose $c_-=c_+=1/2$ and $\kappa= f^{1/3}$) 
\begin{equation}
\label{SW123}
{\cal L } \sim - f^{2/3} \int d^4 \theta \, \Lambda_+ \Lambda_- \overline \Lambda_+ \overline \Lambda_- \, .   
\end{equation}

\subsection{A model with chiral bosons} 

As our next example we study the Lagrangian 
\begin{equation}
\label{GG1}
{\cal L} = \int d^4 \theta \Big{[} - \Sigma \overline \Sigma + \frac{1}{2 f^2} \, \D_+ \S \,  \Db_+ \overline \S \, \Db_- \S \,  \D_- \overline \S  \Big{]} \, . 
\end{equation} 
The bosonic sector of the Lagrangian \eqref{GG1} is given by
\begin{equation}
\begin{aligned}
{\cal L}_{bos.}  =&\ \p_= A \p_{\+} \overline A - P_{\+} \overline{P}_= - \overline{P}_{\+} P_=  -F \overline F + K \overline K + L \overline L 
\\
&
+ \frac{1}{2 f^2} \Big [  F \overline F L \overline L 
- (L \overline L)^2 +  L \overline L P_{\+}(\overline{P}_= + i\p_= \overline A) + L \overline L \overline{P}_{\+}(P_= - i\p_=  A) \Big ] 
\\
&
- \frac{1}{2 f^2}  \, P_{\+} \overline{P}_{\+} (P_= - i\p_= A) (\overline{P}_= + i\p_= \overline A) \,  . 
\end{aligned}
\end{equation}
We now start to integrate out the auxiliary fields. 
It is clear that on top of the trivial supersymmetric solution there are various other solutions to the auxiliary field equations. 
We shall not cover all the possible solutions in detail, 
but we will restrict to one which has some special features, and is easy to handle. 
Taking the variation with respect to the bosonic auxiliary fields 
one can verify the consistent solution 
\begin{equation}
\label{GG2}
\begin{aligned}
F & = K =0 = P_{\+} = P_{=} \, , 
\\
L  & = f \, , 
\\
\partial_{=} A &= \partial_{=} \overline A   = 0 \, .  
\end{aligned}
\end{equation}
We see that the bosonic sector in this solution contains only a chiral scalar, 
and note that the vacuum energy here is negative 
\begin{equation}
\langle V \rangle= - \frac{f^2}{2} \, . 
\end{equation}  
In this background, 
the fermionic sector reads up to quadratic order 
\begin{equation}
{\cal L}_{ferm.}^{(2)}   =   
-i \psi_- \p_{\+} \overline{\psi}_- + i \psi_+ \p_{=} \overline{\psi}_+ 
- \frac{i}{2} \lambda_+ \p_{=} \overline \lambda_+ \, , 
\end{equation}
where we have already integrated out $\lambda_-$ and $\chi_+$ to find that both vanish, 
up to this order. 
Notice that 
\begin{equation}
\begin{aligned}
\langle \delta \psi_- \rangle & = - \epsilon_- f \ ,\\
 \ \langle \delta \lambda_+ \rangle & =  \overline \epsilon_+ f  \, , 
\end{aligned}
\end{equation}
in the background \eqref{GG2} and therefore they are the goldstini. 
From our results in component form we see that in this background the theory contains 
a left-moving scalar $A$ and a left-moving fermion $\psi_+$. 
These together form a left-moving multiplet, which as we will see fits nicely in a chiral lefton. 
In addition the two goldstini fit quite well with the fact that the vacuum energy is non-zero, 
because supersymmetry here is completely broken. 
Moreover, the goldstini have ``wrong'' signs\footnote{ In contrast to bosons a fermionic kinetic term with an opposite sign does not necessarily lead to an ill-defined theory. 
Indeed for the bosons a wrong sign kinetic term would make the free theory (Euclidean) path integral diverge 
whereas fermions with the wrong sign kinetic term does not lead to any divergence due to the properties of Grassmann integration.} for their kinetic terms, 
a property that is aligned with the negative energy as well.

We can verify that the above solution is an exact solution of the full theory, 
by solving the superspace equations of motion, 
which read
\begin{equation}
\label{V41}
\mathrm{D}_{\pm} \left( \Sigma
- \frac{1}{2 f^2} \, \Big \{
\overline{\mathrm{D}}_+ ( \mathrm{D}_-  \overline \Sigma \, \mathrm{D}_+  \Sigma \, \overline{\mathrm{D}}_-  \Sigma)
-\mathrm{D}_- ( \overline{\mathrm{D}}_+  \overline \Sigma \, \mathrm{D}_+  \Sigma \, \overline{\mathrm{D}}_-  \Sigma ) 
\Big \} \right) =  0 \, . 
\end{equation} 
We  see that the equation \eqref{V41} is solved by 
\begin{equation}
\label{SYL}
\Sigma = Y + \overline \Phi_L \, , 
\end{equation} 
where $Y$ is a twisted chiral goldstino superfield, and $\Phi_L$ is the chiral lefton. 
For the lefton we have 
\begin{equation}
\begin{aligned}
\overline{\mathrm{D}}_{\pm} \Phi_L &= 0 \, , 
\\
\mathrm{D}_- \Phi_L  &= 0 \, . 
\end{aligned}
\end{equation} 
Here the chiral lefton serves as the zero mode of equation \eqref{V41}, 
and the solution \eqref{SYL} can clearly exist only in 2D. 
One can show that the twisted chiral goldstino $Y$ solves the superspace equations of motion 
\begin{equation}
\label{V42}
 {Y}
- \frac{1}{2 f^2} \, \Big \{
\overline{\mathrm{D}}_+ ( \mathrm{D}_-  \overline{Y} \, \mathrm{D}_+  {Y} \, \overline{\mathrm{D}}_-  {Y} )
-\mathrm{D}_- ( \overline{\mathrm{D}}_+  \overline{Y} \, \mathrm{D}_+  {Y} \, \overline{\mathrm{D}}_-  {Y} ) 
\Big \}  =  0 \, , 
\end{equation} 
since it satisfies 
\begin{equation}
\label{Y3}
\begin{aligned}
Y^2 &= 0 \, ,  
\\ 
\overline D_- D_+  Y   &= f + 2  \overline {\cal N}  \overline{Y} \, ,  
\end{aligned}
\end{equation} 
where $ \cal N$ is a twisted chiral Lagrange multiplier.

For completeness, it is gratifying to see that here the goldstino sector can be embedded using the Samuel-Wess superfield (\ref{SWdef}) as $\Sigma = \Sigma_L$.

One can find equivalent results by investigating the Lagrangian 
\begin{equation} 
{\cal L} = \int d^4 \theta \Big{[} - \Sigma \overline \Sigma + \frac{1}{2 f^2} \, \D_- \S \,  \Db_- \overline \S \, \Db_+ \S \,  \D_+ \overline \S  \Big{]} \, ,  
\end{equation} 
but now it is the auxiliary field $K$ which will get a {\it vev}, 
and we will have the full superspace solution given by 
\begin{equation}
\label{SYR}
\Sigma = \overline Y + \overline \Phi_R  \, . 
\end{equation}
The low energy model for this setup will be described by $\Sigma=\Sigma_K$.

\subsection{A model containing only goldstini}

In the previous examples we studied models with additional propagating 
non-goldstino multiplets in the broken vacuum. 
This is not essential, 
and we can illustrate this with a simple example. 
Consider the Lagrangian 
\begin{equation} 
\label{V11} 
{\cal L} =  \int d^4 \theta \Big{[}  \overline{\S }^2 + \S^2  
- \frac{2}{f^2} \, \D_+ \overline \S \, \D_- \overline \S \,  \Db_+ \S \, \Db_- \S \Big{]} \, . 
\end{equation} 
As usual, our strategy to find the new vacua is to first write down the bosonic sector, 
and then investigate the existence of non-trivial solutions to the auxiliary field equations. 
The bosonic sector reads 
\begin{equation} 
\begin{aligned}
\label{V12} 
{\cal L}_{bos.} =& - 2 K L - 2 \overline K \overline L 
+2 (P_= - i \p_= A)(P_{\+} - i \p_{\+} A) 
+2 (\overline P_= + i \p_= \overline A)(\overline P_{\+} + i \p_{\+} \overline A)
\\
& - \frac{2}{f^2}  \Big{[} 
\overline K \overline L (P_{\+} - i\p_{\+}A)(P_= - i\p_= A) 
+ K L (\overline{P}_{\+} + i\p_{\+}\overline A) (\overline{P}_= + i\p_= \overline A) \Big{]} 
\\
& + \frac{2}{f^2}  \Big{[} (P_{\+} - i\p_{\+}A)  (P_= - i\p_= A) (\overline{P}_{\+} + i\p_{\+}\overline A)(\overline{P}_= + i\p_= \overline A)   + K \overline K L \overline L \Big{]} \, .  
\end{aligned}
\end{equation} 
From the variation with respect to the vector auxiliary fields we have 
\be
P_= =  i\p_= A  \ , \ P_{\+} = i\p_{\+}A  \, , 
\ee
which, as will become clear later, means that the scalar $A$ is not propagating, 
since it drops out of the theory. 
From the variation with respect to the auxiliary fields $K$ and $L$ we find 
\begin{equation} 
\begin{aligned} 
0 = & L \left( 1 - \frac{1}{f^2}  \, \overline K \overline L \right) \, , 
\\
0 = & K \left( 1 - \frac{1}{f^2}  \, \overline K \overline L \right) \, , 
\end{aligned} 
\end{equation} 
which have the trivial solution $K=L=0$, 
but they also have the solution 
\begin{equation}
\label{V33}
\begin{aligned}
L = f  \ , \ 
K  = f  \, . 
\end{aligned}
\end{equation} 
In this background the complete Lagrangian up to quadratic order in fermions becomes 
\begin{equation}
\label{V34}
{\cal L}^{(2)} = - 2 f^2  + i  \psi_+ \p_{=} \overline{\psi}_+ + i  \psi_- \p_{\+} \overline \psi_-  \, . 
\end{equation} 
From \eqref{V34} we see that there are no propagating bosons in the theory, 
but there are the propagating fermions $\psi_+$ and $\psi_-$. 
Due to the presence of the positive vacuum energy in \eqref{V34} we know that supersymmetry is broken, 
and we can identify the goldstini 
\begin{equation}
\begin{aligned}
\label{GGYY}
\langle \delta \psi_+ \rangle & = \epsilon_+  f\, , 
\\
\langle \delta \psi_- \rangle & = - \epsilon_- f\,  . 
\end{aligned}
\end{equation}

We can now verify that this is a consistent solution to the full component equations of motion. 
A straightforward way to see this is by finding the complete equivalent solution to the superspace equations of motion. 
The variation with respect to $\overline \S$ of the superspace Lagrangian reads 
\begin{equation}
\mathrm{D}_\pm \Big\{ \overline \Sigma
+ \frac{1}{f^2} \big [
 \mathrm{D}_+ ( \mathrm{D}_-  \overline \Sigma  \overline{\mathrm{D}}_+  \Sigma \overline{\mathrm{D}}_-  \Sigma  ) 
- \mathrm{D}_- ( \mathrm{D}_+  \overline \Sigma  \overline{\mathrm{D}}_+  \Sigma \overline{\mathrm{D}}_-  \Sigma  )
\big ] \Big \} = 0 \, .
\end{equation}
This equation is solved by 
\begin{equation}
\label{SYY}
\S = Y + \overline Y \, ,  
\end{equation}
where $Y$  satisfies \eqref{Y1} and \eqref{Y2}. 
The solution \eqref{SYY} describes exactly the two on-shell goldstini degrees of freedom \eqref{GGYY}. 
One can now find the full solution for each single component field of the multiplet $\Sigma$, 
by using appropriate projections of the superspace solution \eqref{SYY}.

Also here the Goldstino sector can be embedded using the Samuel-Wess superfield through the combination $\Sigma_L + \Sigma_K$.

\section{Models with semichiral multiplets in $\mathbf{(2,2)}$}

In the previous section we studied examples with complex linear superfields. 
To illustrate the generality of the discussion, 
we give an example with a left semichiral multiplet which exists only in 2D and has no equivalent in 4D.

We have the Lagrangian 
\begin{equation}
{\cal L} =  \int d^4 \theta \Big{[}  - \XL  \XLb  + \frac{1}{2f^2} \,  \D_+ \XL \Db_+ \XLb \Db_- \XL \D_- \XLb   \Big{]} \, . 
\end{equation} 
To find the possible backgrounds for this model we start from the bosonic sector 
\be
\begin{aligned}
{\cal L}_{bos.} = & \partial_= A \partial_\+ \overline A  
- i \partial_\+ A \, \overline P_=  
+ i \partial_\+ \overline A \, P_= 
-F \overline F + L \overline L 
\\
& + \frac{1}{2f^2}  \, \Big{[}   F \overline F \, L \overline L - L^2 \overline L^2 
+i \, L \overline L \, \partial_\+ A \, (\overline P_= + i \partial_= \overline A )  
-i \, L \overline L \,  \partial_\+ \overline A \, ( P_= - i \partial_=  A )   
\Big{]} 
\\
& - \frac{1}{2f^2}  \, \partial_\+ A \, \partial_\+ \overline A  \, (P_= - i \partial_=  A) (\overline P_= + i \partial_= \overline A ) \, . 
\end{aligned}
\ee
To integrate out the auxiliary fields, we have to solve their equations of motion.
We choose the particular solution
\be
\label{RR}
\begin{aligned}
F &  = 0 \, , 
\\
L &  = f \, , 
\\
P_= & = 0 \, , 
\\
\partial_\+ A & = 0 \, .   
\end{aligned}
\ee
{}From this we see that the theory contains a right moving scalar $A$. 
In addition the vacuum energy of the theory is negative $\langle V \rangle = - \frac{f^2}{2}$. 
We will see that this righton forms a supermultiplet with a right moving fermion. 
In particular, once we write the fermionic sector in the background \eqref{RR} we have for the quadratic fermionic terms 
(after we integrate out the fermion auxiliary fields) 
\be
{\cal L}_{ferm.}^{(2)} = - i  \psi_- \partial_\+ \overline \psi_-    
- \frac{i}{2}  \lambda_{+} \partial_= \overline \lambda_{+}      
- i  \lambda_{-} \partial_\+ \overline \lambda_{-}  \, . 
\ee 
Notice that 
\begin{equation}
\begin{aligned}
\langle \delta \lambda_{+} \rangle & = f \overline \epsilon_+ \, , 
\\
\langle \delta \psi_- \rangle & =  - f \epsilon_- \, . 
\end{aligned}
\end{equation} 
Therefore the fermion superpartner of the right moving scalar is the right moving fermion $\lambda_-$, 
whereas $\lambda_{+}$ and $\psi_-$ are the goldstini.

As in our previous examples, 
we can verify that this is a consistent background because we can solve the superspace equations of motion exactly. 
In particular the superspace equations of motion read 
\be
\label{whynot}
\XLb  
- \frac{1}{2f^2}  \,  \D_+ ( \Db_+ \XLb \Db_- \XL \D_- \XLb ) 
+ \frac{1}{2f^2}  \, \Db_- (  \D_+ \XL \Db_+ \XLb  \D_- \XLb ) =  \mathbb{Y}_L \, ,  
\ee
where we have introduced the left semichiral zero-mode 
\be
\Db_+ \mathbb{Y}_L = 0 \, . 
\ee 
To decouple the equations for the zero-mode from the equations of the supersymmetry breaking sector we find the 
constraints\footnote{ After introducing $\mathbb{Y}_L$ the consistency condition is $ \tfrac{1}{2f^2}  \, \D_+ \Db_- (  \D_+ \XL \Db_+ \XLb  \D_- \XLb ) = D_+  \mathbb{Y}_L$. If $\D_+ \mathbb{Y}_L = 0$ then the zero-mode does not depend on the breaking sector and one needs to check the consistency of the solution only for the goldstino part. } 
\be
\D_+ \mathbb{Y}_L = 0 = \D_- \mathbb{Y}_L \, , 
\ee
which make $\mathbb{Y}_L $ a chiral righton, 
which is consistent with our results from the component form. 
For the supersymmetry breaking sector we can easily see that the twisted chiral superfield $Y$ in \eqref{twistednilp} 
solves the equations when we ignore the zero-mode. 
Eventually, 
the solution to the full equations of motion \eqref{whynot} is given by 
\be
\XLb = \overline Y  + \mathbb{Y}_L  \, . 
\ee

One can also construct models with chiral superfields (also real linear in 4D) including superspace higher derivative terms. 
These models have been extensively studies in the literature 
\cite{
Cecotti:1986jy,
Antoniadis:2007xc,
Khoury:2010gb,
Koehn:2012ar,
Farakos:2012qu,
Nitta:2014fca,
Aoki:2014pna,
Dudas:2015vka,
Aoki:2015eba,
Ciupke:2015msa,
Bielleman:2016grv,
Aoki:2016cnw,
Ciupke:2016agp,
Fujimori:2016udq,
Bielleman:2016olv} 
and we refer the reader to these publications for further properties.

\section{Higher derivatives and supersymmetry breaking in $\mathbf{(4,4)}$}

In this section we will discuss a model with $(4,4)$ supersymmetry. 
The model we will study can be built also in  4D $N=2$. 
The $(4,4)$ superspace derivatives satisfy the algebra 
\be
\begin{split}
\{ \D_- , \Db_- \} &= i \p_=  \, , 
\\
\{ \D_+ , \Db_+ \} &= i \p_\+  \,  , 
\\
\{ \nabla_- , \overline{\nabla}_- \} &= i \p_=  \,  , 
\\
\{ \nabla_+ , \overline{\nabla}_+ \} &=  i \p_\+ \,  , 
\end{split}
\ee
whereas all other super-commutators vanish. 
The $(4,4)$ has various  multiplets, 
but here we will utilize an unconstrained complex superfield, 
which is sufficient for our study. One can eventually start to build more realistic models either in 2D $(4,4)$ 
or in the equivalent theory of 4D $N=2$. 
In any case, the simple model we give here will be underlying any more realistic extension.  
Thus we have 
\be
Z : \text{unconstrained $(4,4)$ superfield.} 
\ee
The various component fields of the $Z$ superfield are defined as usual by projections. 
In particular we will use the component fields 
\be
\begin{split}
& \D^2 \nabla^2 Z | = F \, , 
\\
& \nabla_\alpha \D^2 Z |= \psi_\alpha \, , 
\\
& \D_\alpha \nabla^2 Z |=  \lambda_\alpha  \, , 
\\
& \overline \nabla_{\dot \alpha} \D^2 \nabla^2 Z | = \overline \chi_{\dot \alpha} \, , 
\\
&\overline \D_{\dot \alpha} \D^2 \nabla^2 Z | = \overline \sigma_{\dot \alpha}   \, , 
\end{split}
\ee
where $\nabla^2= \nabla_+ \nabla_-$ and so on. 
Clearly, 
if one considers the free theory 
\be
\label{Zfree}
{\cal L} = - \int d^8 \theta Z \overline Z \, , 
\ee
all the components will work as Lagrange multipliers which will  set one another to vanish. 
In other words, 
all the fields in the theory defined by \eqref{Zfree} are auxiliary. 
This is most easily seen by looking at the superspace equations of motion of the free model \eqref{Zfree} which read 
\be
Z =0 \, , 
\ee
because $Z$ is unconstrained. 
We will see that once we deform the theory by introducing higher derivatives, 
there will be new vacua where the auxiliary fields pick up kinetic terms and start to propagate.

The Lagrangian reads 
\be
\label{motorhead}
\begin{split}
{\cal L} =& -  \int d^8 \, \theta  Z \overline Z  
\\
&+ \frac{1}{4f^6} \int d^8 \theta \, 
\D_+ \nabla^2 Z \, \D_- \nabla^2 Z  \, 
\Db_+ \overline \nabla^2 \overline Z \, \Db_- \overline \nabla^2 \overline Z  \, 
\nabla_+ \D^2 Z \, \nabla_- \D^2 Z \, 
\overline \nabla_+ \Db^2 \overline Z \, \overline \nabla_- \Db^2 \overline Z \, . 
\end{split}
\ee
Here $f$ is a constant related to the supersymmetry breaking scale as we will see. 
Once we expand \eqref{motorhead} in components we find 
\begin{equation} 
\begin{aligned}
\label{V52} 
{\cal L} =& - F \overline F + \frac{1}{4 f^6} (F \overline F)^4 
\\
& +  \left( -1
+ \frac{(F \overline F)^3}{f^6}\right) \left [ \sigma_+ \lambda_- + \lambda_+ \sigma_- + \chi_+ \psi_- + \psi_+ \chi_- + c.c. 
\right ]
\\
& + \left ( 1 - \frac{(F \overline F)^3}{4 f^6}\right ) 
\left [ i \overline \lambda_+ \p_= \lambda_+  + i \overline \lambda_- \p_{\+} \lambda_-
+  i \overline \psi_+ \p_= \psi_+  + i \overline \psi_- \p_{\+} \psi_- \right ] + \dots \, ,  
\end{aligned}
\end{equation} 
where the $\dots$ contains terms which are not essential for our discussion. 
Notice that 
\be
\begin{aligned}
\delta \lambda_\alpha &= \epsilon_\alpha F + \cdots \, ,
\\
\delta \psi_\alpha &= \eta_\alpha F + \cdots \, ,
\end{aligned}
\ee
where $\epsilon_\alpha$ and $\eta_\alpha$ are the supersymmetry parameters of $(4,4)$, 
therefore the fermions  $\psi_\alpha$ and $\lambda_\alpha$ will be goldstini when $\langle F \rangle \ne 0$.

We first want to find the new backgrounds this theory can have. 
From the structure of the Lagrangian it is consistent to set all bosonic fields to vanish, 
while we keep the field $F$ non-vanishing. 
We ask the reader to bear with us at this point and 
later we will explicitly solve the complete equations of motion thereby justifying our assumption. 
Once we set all the other bosons to zero except $F$, the bosonic sector will reduce to 
\be
\label{LF}
{\cal L}_F = -  F \overline F 
+ \frac{1}{4 f^6}  F^4 \overline F^4 , 
\ee
with equations of motion for $F$ 
\be
\overline F = \frac{1}{f^6} \, \overline F \, (F \overline F)^3 . 
\ee
This equation has two solutions. 
First there is the supersymmetric solution $F=0$, which corresponds to $Z=0$. 
Second we can have 
\be
\label{FF}
F  = f \, ,
\ee
which breaks supersymmetry. 
Notice that \eqref{FF} gives only the {\it vev} of $F$ not the complete solution, 
which we will show how it can be found when we turn to superspace. 
For the vacuum energy we find by inserting solution \eqref{FF} into \eqref{LF} that 
\be
\langle V \rangle =  \frac{3f^2}{4} \,  . 
\ee
Therefore supersymmetry is broken with positive  vacuum energy. 
In addition to this, 
once we expand the model \eqref{motorhead} in the background  \eqref{FF}  we see it gives rise to kinetic terms for the 
goldstini as expected  
\be
\label{Ffermvac}
{\cal L} = - \frac{3 f^2}{4} 
+\frac{3 i}{4} \overline \psi_+ \p_= \psi_+
+\frac{3 i}{4} \overline \psi_- \p_{\+} \psi_- 
+\frac{3 i}{4} \overline \lambda_+ \p_= \lambda_+
+\frac{3 i}{4} \overline \lambda_- \p_{\+} \lambda_-    
+ \cdots \, ,
\ee
The terms in $\cdots$ will contain higher order goldstini self-interactions which make the theory invariant under $(4,4)$ non-linearly realized supersymmetry \cite{Cribiori:2016hdz}. 
It is gratifying to notice that the dangerous terms in the second line of \eqref{V52} vanish in this background.

Now we prove that the solution \eqref{FF} can be embedded into a complete superspace solution, 
which describes a theory where $(4,4)$ is spontaneously broken and non-linearly realized. 
To this end we rapidly review the goldstino model for 2D  $(4,4)$, 
which is in fact a dimensional reduction of the 4D $N=2$. 
Following \cite{Kuzenko:2011ya,Cribiori:2016hdz}, the superfield which one has to employ is 
a chiral $(4,4)$ superfield $X$ 
\be
\Db_{\alpha} X =0 \ , \ \overline \nabla_\alpha X = 0 \,  , 
\ee
which satisfies the constraints 
\be
\label{C} 
X \, \nabla_\alpha \D^2 \, X = 0 \ , \ X \, \D_\alpha \nabla^2 \, X = 0 \, . 
\ee
The superspace solution to the constraints \eqref{C} is 
\be
\label{CCC}
X = \frac{\D_+  \nabla^2  X \,  
\D_-  \nabla^2  X  \, 
\nabla_+ \D^2  X \, 
\nabla_- \D^2  X}{(\nabla^2 \D^2  X)^3} \, . 
\ee
The only independent components inside $X$ are defined by the projections  
\be 
\D^2 \nabla^2 X | = {\cal F} \  , \  \nabla_\alpha \D^2 X |= \tilde g_\alpha \ , \  \D_\alpha \nabla^2 X |=  g_\alpha  \, .  
\ee
Here $g_\alpha$ and $\tilde g_\alpha$ are the goldstini and ${\cal F}$ is the auxiliary field which gets a {\it vev} and breaks supersymmetry. 
Incidentally, 
due to \eqref{CCC}, $X$ also satisfies 
\be
X^2 =0 \, . 
\ee 
The full expansion of $X$ can be found in \cite{Cribiori:2016hdz}. 
The minimal Lagrangian for $X$ is 
\be
\label{tc}
{\cal L} = \int d^8 \theta \, X \overline X - \left \{ f \int d^4 \theta^c \, X + c.c. \right \} \, , 
\ee
where $\theta^c$ refers to the chiral part of the full $(4,4)$ superspace. 
Here we want to find the superspace equations of motion that arise from \eqref{tc}. 
To achieve this we start from the Lagrangian 
\be
\label{LXC}
\begin{aligned}
{\cal L} = & \int d^8 \theta  \left \{ X \overline X 
+ \left ( {\cal C}^\alpha \, X  \, \D_\alpha \nabla^2 \, X 
+ \tilde {\cal C}^\alpha \, X \, \nabla_\alpha \D^2 \,X + c.c. \right )
\right \} \\
&-  f \left( \int d^4  \theta^c X +  c.c. \right) \, , 
\end{aligned}
\ee
where now $X$ is an $(4,4)$ chiral superfield but otherwise unconstrained, 
and ${\cal C}^\alpha$ and $\tilde {\cal C}^\alpha$ are Lagrange multiplier superfields 
which will impose the constraints \eqref{C}. 
Indeed, once we perform the superspace variation we find the equations \eqref{C} along with 
\be
\label{FFeq}
X \, \Db^2 \overline \nabla^2 \overline X &=& f X \, , 
\\
\label{Ga}
G^2 \,  \tilde G_\alpha \, \overline X \, \Db^2 \overline \nabla^2 \overline X &=& f \overline X \, G^2 \,  \tilde G_\alpha \, , 
\\
\label{Gb}
\tilde G^2 \,  G_\alpha \, \overline X \, \Db^2 \overline \nabla^2 \overline X &=& f \overline X \, \tilde G^2 \,  G_\alpha \, . 
\ee
In \eqref{Ga} and in \eqref{Gb} we use $G_\alpha$ to represent the superfield $\D_\alpha \nabla^2 X$, 
which has the goldstino $g_\alpha$ in the lowest component. 
Similarly $\tilde G_\alpha = \nabla_\alpha \D^2 X$. 
The physical meaning of these equations is the following: 
Equation \eqref{FFeq} gives the equation which once solved gives the on-shell value for ${\cal F}$, 
and equations \eqref{Ga} and \eqref{Gb} are in fact equivalent to the equations of motion for the goldstini $g_\alpha$ 
and $\tilde g_\alpha$.

Now we return to the model \eqref{motorhead} and we are ready to complete the superspace solution that describes the 
vacuum where supersymmetry is spontaneously broken. 
The superspace equations of motion which follow from \eqref{motorhead}  are 
\be
\nn
\overline Z = \frac{1}{4f^6} \Big{[} \!\!\! && -\nabla^2 \D_+ \left( \D_-  \nabla^2 Z \, \nabla_+ \D^2 Z \, \nabla_- \D^2 Z \, \overline {\cal B}  \right) 
\\
\nn
&& +  \nabla^2 \D_- \left( \D_+  \nabla^2 Z \, \nabla_+ \D^2 Z \, \nabla_- \D^2 Z \, \overline {\cal B}  \right) 
\\
\nn
&& - \D^2 \nabla_+ \left(  \D_+ \nabla^2 Z \, \D_- \nabla^2 Z \, \nabla_-  \D^2 Z \, \overline {\cal B}  \right)  
\\
\label{EOMZ}
&& + \D^2 \nabla_- \left( \D_+ \nabla^2 Z \, \D_- \nabla^2 Z \, \nabla_+  \D^2 Z \,  \overline {\cal B}  \right)   \Big{]} , 
\ee
where 
\be
\overline {\cal B} = 
\Db_+ \overline \nabla^2 \overline Z \,  
\Db_- \overline \nabla^2 \overline Z  \, 
\overline \nabla_+ \Db^2 \overline Z \, 
\overline \nabla_- \Db^2 \overline Z . 
\ee
It is now straightforward to verify  that 
\be
\label{ZX}
Z = X 
\ee
is  an exact solution  to the equation \eqref{EOMZ}. 
The $X$ in \eqref{ZX} satisfies \eqref{FFeq}, \eqref{Ga}, \eqref{Gb} and \eqref{C}.  
Therefore the goldstini inside $X$ are {\it on-shell}. 
This is expected, because the equation \eqref{EOMZ}  is as we said an equation of motion, 
therefore describes also equations of motion for the dynamical fields. 
This concludes the proof that the solution \eqref{FF} is consistent, 
since we have found it has a full extension in superspace given by \eqref{ZX}. 
From here one can find all the on-shell values of the other fields inside $Z$. 
Finally, 
the theory in the broken vacuum can be effectively represented by a Lagrangian of the form \eqref{LXC}, 
since it has to give rise to the same equations of motion. 
Other forms of this Lagrangian can be found in \cite{Cribiori:2016hdz}. 
Different superspace methods for the effective description of broken $N=2$ supersymmetry in 4D can be found in \cite{Kuzenko:2011ya}.

Of course the equation \eqref{EOMZ} has  also a second solution, the standard solution, 
which is  
\be
Z=0  , 
\ee
which describes the original supersymmetric vacuum. 
We have seen again that superspace higher derivatives change the degrees of freedom in the theory while they give access to new vacua.

\section{Conclusions and prospects}

Superspace higher derivative terms have applications ranging from supersymmetry breaking to inflationary cosmology and only recently they have started to receive more attention. 
In this work we focused on two-dimensional models and we have initiated a systematic study of their properties, 
and in particular we focused on finding new vacua. 
The 2D theories with higher derivative terms have a richer vacuum structure than their 4D relatives. This is due to the fact that the 2D nonminimal multiplets have more auxiliary fields but also that there are more possible terms that can be written down in 2D which would not be Lorentz invariant in 4D. Furthermore, in 2D there are several different nonminimal multiplets that can be used. 
The examples we have presented in this paper were chosen because of their simplicity and because they illustrate some typical features of the supersymmetry breaking vacua that can occur. 
Clearly there are many more solutions of the models presented and many more models that may be studied. 
In general it seems that the additional solutions which do not exist in four dimensions tend to be more complicated with exotic particle content including for instance chiral bosons.
Some of the models we have presented here have the additional property to be equivalent to 4D $N=1$ and 4D $N=2$ theories. 
The 4D $N=2$ models we have discussed are presented here for the first time and offer novel methods to break $N=2$ 
supersymmetry down to $N=0$. 
In addition, these models give directly a non-linear realization of supersymmetry. 
We believe that the extension to higher $N$ is straightforward, 
but embedding this mechanism into realistic theories might present a challenge which we leave for future work.  
Finally, one may wonder about the possibility of perturbatively generating the class of higher derivative terms considered in this paper. This direction was investigated in \cite{Farakos:2014iwa} where it was shown that although these terms do appear in the one-loop effective action, they are generically accompanied by other terms that ruin the symmetry breaking effect. This nicely corresponds to the fact that in the limit $f\rightarrow \infty$ the symmetry breaking vacuum becomes ill-defined. About a possible nonperturbative origin of these terms, we can only speculate.

\acknowledgments 

We thank N. Cribiori, O. Hul\'{i}k, K. Koutrolikos, U. Lindstr\"om and M. Ro\v{c}ek for discussions. 
P.K. would like to thank the University of Padova for the kind hospitality during the preparation of this work. 
This work is supported in part by the MIUR grant RBFR10QS5J (STaFI) and by the Padova University Project CPDA119349. 
This work is supported by the Grant agency of the Czech republic under the grant P201/12/G028.

\appendix

\section{Scalar multiplets in 2D}

An intrinsic property of supersymmetric theories is that they contain auxiliary fields. 
These fields are useful in closing the supersymmetry algebra, 
and are eventually integrated out in the component form level, 
so that we find the theory in terms of the propagating modes. 
The interesting feature of supersymmetric multiplets is that one can dress a scalar multiplet (which contains a scalar and its fermion superpartner) 
with a large diversity of auxiliary fields which are bosonic and fermionic. 
In this section we briefly remind the reader the various possibilities.

The less restrictive structure of supersymmetry in two dimensions is reflected in the number of possible multiplets: 
\begin{multicols}{2}
\begin{itemize}
\item The chiral superfield:\\
$\overline{\D}_{\pm} \Phi = 0 = \D_{\pm} \overline \Phi \, .$
\item The twisted chiral superfield:\\ 
 $\overline{\D}_+ \chi = \D_- \chi= 0 = \D_+ \overline \chi = \overline{\D}_- \overline{\chi} \, .$
\item The complex linear superfield:\\ 
$\overline{\D}_+ \overline{\D}_- \S = 0 = \D_+ \D_- \Sb \, .$
\item The twisted complex linear superfield:\\
 $\overline{\D}_+ \D_- \Omega = 0 = \D_+ \overline{\D}_- \overline \Omega \, .$
\item The left semichiral superfield:\\
 $\overline{\D}_+ \XL  = 0 = \D_+ \overline{\mathbb{X}}_L \, .$
\item The right semichiral superfield:\\
 $\overline{\D}_- \XR  = 0 = \D_- \overline{\mathbb{X}}_R \, .$ 
\end{itemize}
\end{multicols}
A chiral superfield contains a complex scalar, two fermions and a complex auxiliary field. In contrast to a chiral superfield,  a complex linear superfield contains a bigger set of auxiliary fields. However, if the auxiliary fields of the free theory do not become propagating, both of these superfields offer the same on-shell description. A semichiral superfield consists of twice as many components as the chiral superfield. To define a consistent free theory, semichiral superfields need to come in pairs $(\XL,\XR)$. The simplest consistent action is given by 
\be
{\cal L} &=& -\int d^4 \theta \  \XLb \XR + c.c.
\ee 
There are also twisted chiral superfields and twisted complex linear superfields, 
which of course have the same number of components as their untwisted cousins. Moreover, in the presence of particular symmetries it can be shown that there is a duality between chiral and twisted chiral superfields whereas the semichiral doublet can be dualized to a chiral and a twisted chiral superfield \cite{Grisaru:1997ep}.

The components of a chiral superfield $\Phi$ are defined by 
\be
\Phi | \equiv A \, ,
\, \D_{\pm} \Phi | \equiv \lambda_{\pm} \, ,
\,\D_{+} \D_{-} \Phi | \equiv  F \, . 
\ee 
A complex linear superfield $\S$ has components 
\begin{align}
\S| &\equiv   A \, , &\overline{\D}_{+} \D_{+} \S| &\equiv   P
_{\+} \, , \nonumber\\
\D_{\pm} \S | &\equiv \lambda_{\pm} \, ,
&\overline{\D}_{-} \D_{-} \S| &\equiv   P_{=} \, , \nonumber\\
\overline{\D}_{\pm} \S | &\equiv \psi_{\pm} \, ,
&\overline{\D}_{-} \D_{+}\S| &\equiv   L \, , \\
\D_{+} \D_{-} \S| &\equiv   F \, , &\frac{1}{2}\left( \D_{+} \overline{\D}_{\pm} \D_{-} - \D_{-} \overline{\D}_{\pm} \D_{+}\right)\S| &\equiv  \s_{\pm} \, ,\nonumber\\
\overline{\D}_{+} \D_{-} \S| &\equiv   K \, .\nonumber
\end{align}  
The general supersymmetry transformations of the fermions $\lambda, \psi$ in the complex linear superfield are 
\begin{align}
\delta \lambda_{+} &= \epsilon_{+}F + \overline{\epsilon}_{+}L - \overline{\epsilon}_-P_{\+} \, ,
\nonumber \\
\delta \lambda_{-} &= \epsilon_{-}F - \overline{\epsilon}_{-}K + \overline{\epsilon}_+P_{=} \, ,
\nonumber \\
\delta \psi_{+} &= \epsilon_{+}K - \epsilon_{-}P_{\+} +i \epsilon_- \p_{\+}A \, ,\\
\delta \psi_{-} &= -\epsilon_{-}L + \epsilon_{+}P_{=} -i \epsilon_+ \p_{=}A \, .
\nonumber
\end{align} 
The components of a semichiral superfield $\XL$ are defined by 
\begin{align}
\XL| &\equiv   A \, , &\overline{\D}_{-} \D_{-} \XL| &\equiv   P
_{=} \, , \nonumber\\
\D_{\pm} \XL | &\equiv \lambda_{\pm} \, ,
&\overline{\D}_{-} \D_{+}\XL| &\equiv   L \, ,\nonumber\\
\overline{\D}_{-} \XL | &\equiv \psi_{-} \, ,
&\frac{1}{2}\left( \D_{+} \overline{\D}_{-} \D_{-} - \D_{-} \overline{\D}_{-} \D_{+}\right)\XL| &\equiv  \s_{-} \, , \\
\D_{+} \D_{-} \XL| &\equiv   F \,  \nonumber.
\end{align} 
Spinorial measures are defined as 
\begin{equation} 
\begin{aligned}
\int d^4 \theta \ {\cal K} &= -  \mathrm{D}_+  \mathrm{D}_- \overline{\mathrm{D}}_+    \overline{\mathrm{D}}_- \  {\cal K} \, , 
&\int d^8 \theta \ {\cal K} &=   \mathrm{D}_+  \mathrm{D}_- \overline{\mathrm{D}}_+    \overline{\mathrm{D}}_- \  
\nabla_+ \nabla_-
\overline \nabla_+ \overline \nabla_-
{\cal K} \, , \\
\int d^2 \theta \ {\cal K} &=   \mathrm{D}_+  \mathrm{D}_- \  {\cal K} \, , \ 
&\int d^2 \theta^T \ {\cal K} &=   \mathrm{D}_+  \overline{\mathrm{D}}_- \  {\cal K} \, , \\
\int d^4 \theta^c \ {\cal K} &=   \D_+  \D_- \nabla_+   \nabla_- \  {\cal K} \, .   
\end{aligned}
\end{equation}

\end{document}